\documentstyle[12pt]{article}

\begin{document}

dedicated to V.I.Arnold's 60-th birthday

\vspace{0.5cm}
\centerline{\LARGE S.P.Novikov\footnote {University
 of Maryland at College Park,
College Park, Maryland 20742-2431 and Landau Institute for
Theoretical Physics,
Kosygin str 2, Moscow 117940, e-mail novikov@ipst.umd.edu.
Research supported in part by the NSF Grant DMS9704613}}

\vspace{0.3cm}
\centerline{\Large Schrodinger Operators on Graphs}
\centerline{and}
\centerline{\Large  Symplectic Geometry\footnote {This work was submitted to
the Fields Institute by the February 28 1998. After the very fruitful
discussions with D.Kazdan, B.Mityagin and P.Kuchment in March-May 1998
several corrections have been made: some definitions were improved,
concrete specific examples were included in the more General
 Examples 2 and 3 below;
 Remark was added to the text of Appendix
concerning  the continuous operators
on graphs;
 the text of Part 2 of the Appendix was improved.}}

\vspace{0.4cm}

{\bf Introduction}. Hamiltonian Formalism of Analytical Mechanics
has been systematically used after Poincare especially by people
who created Quantum Mechanics in the 20's. In pure mathematics,
formalism of differential forms appeared as a  by-product
of Hamiltonian Theory formalized finally by E.Cartan.
 However, geometrical understanding of many important parts
  of Hamiltonian Formalism 
 has not been elaborated on for a long period. For example,
 general definitions of such fundamental geometrical objects
as Lagrangian Submanifolds in  Symplectic (Hamiltonian)
linear spaces (and in more general nonlinear symplectic manifolds
as well) were finally formulated  only in the 60's.
 In particular, V.Arnold
participated in this (see \cite{AA,AF}).

In the late 60's, the present author observed that some algebraic version
of Hamiltonian Formalism
for rings with involutions
plays a fundamental role in many constructions of Differential
Topology (see \cite{N70}). As a reaction to this work, I.M.Gelfand
observed (and pointed out to me in about 1971)
 that von Neumann's construction of selfadjoint
extensions for  symmetric operators is based in fact on some
 Lagrangian planes
in   Hilbert spaces with symplectic scalar product.

The present work is in a sense a natural continuation of these
old observations. An essential part of its results already was announced
by the present author
in a short note \cite{N97}.Let me  point out here that we started
at first to discuss graphs with A.Veselov as a continuation of
a series of papers dedicated to factorization and Laplace transformations
of the Schrodinger operators (see \cite{NV97,ND97}). 

\vspace{1cm}

\centerline{\Large Discrete Schrodinger Operators on Graphs}
\vspace{0.2cm}

\centerline{\Large Wronskians and Topology}
\vspace{0.3cm}

We shall consider Graph $\Gamma$--i.e., one dimensional simplicial complex
such that:

1. Only a finite number of edges $R_i$ (equal to $m_P$)
can meet each other in any vertex $P$;

2.Graph $\Gamma$ has no ends, i.e. $m_P>1$ for every vertex $P$.
Two spaces of scalar complex--valued functions will be used:

\noindent The space of complex-valued functions $\psi_P$
depending on vertices $P$
and the space of functions $\psi_R$ depending on edges $R$.
We do not formulate any global
 restrictions on these functions now.

We shall consider  real self-adjoint (i.e. symmetric, in fact)
operators $L$ acting on these spaces of functions:

\begin{eqnarray}
(L\psi)_P=\sum_{P:P'}b_{P:P'}\psi_{P'}, b_{P:P'}=b_{P':P}\\
(L\psi)_R=\sum_{R:R'}d_{R:R'}\psi_{R'}, d_{R:R'}=d_{R':R}
\end{eqnarray}

Reality means that all coefficients are real.

\newtheorem{deff}{Definition}

\begin{deff}

  Operator $L$ will be called {\bf Finite Type} iff  for any point $P$
 or edge $R$ there exists only a finite number of points $P'$ or edges
 $R'$ such that the coefficients above are nonzero. The Operator
  will be called
 an operator of {\bf Finite Order} iff this number does not depend
  on $P$ or $R$.
 {\bf Order of operator} is a maximal number of vertices (edges) in the
 minimal paths
 joining such pairs of vertices (edges) for which coefficients are nonzero.
 For the {\bf Second Order Operators} on the vertices nonzero coefficients
 can be only $b_{P:P}=V_P$ and $b_{P:P'}$ iff $P\bigcup P'=\partial R$.
 In the case of edges nonzero coefficients can be only $d_{R:R}=V_R$
 and $d_{R:R'}$ iff $R\bigcap R'\neq \emptyset$. We call coefficients
  $V_P$ and $V_R$ {\bf Potentials}

  \end{deff}

For any simplicial complex $K$ and number $k$
we have a boundary operator $\partial$ from
 $k$-chains into $k-1$-chains. We also have  a
 scalar product where $\delta$-functions (i.e. functions
 taking value 1 on one of
 $k$-simplices only) give
an orthonormal basis in subspace of finite chains (i.e. finite functions from
$k$-simplices).
 Therefore we have an adjoint
coboundary operator $\partial^{\star}$.

  Combinatorial {\bf Laplace-Beltrami
Operators} $\Delta_k$ on the spaces of $k$-chains are defined by
the general formula
$$\Delta_k=(\partial+\partial^{\star})^2=\partial\partial^{\star}+
\partial^{\star}\partial$$

For the cases $k=0$ and $k=1$, Laplace-Beltrami operators on  Graph
$\Gamma$ are the second order operators (Schrodinger Operators on
the vertices and edges) such that

\begin{eqnarray}
\Delta_0: b_{P:P'}=1, P\bigcup P'=\partial R, V_P=-m_P\\
\Delta_1: d_{R:R'}=1, R\bigcap R'\neq \emptyset, V_R=-2
\end{eqnarray}

\newtheorem{exa}{Example}
\begin{exa}

Let Graph $\Gamma$ is a line $R^1$, i.e., 1-simplices $R$ and vertices $P$
can be numerated naturally by the integers $n$ such that
boundary of the edge with number $n$ is equal to the union of vertices with
numbers $n$ and $n-1$. We always have $m_P=2$ here. There is a natural
one--to--one correspondance here between edges and vertices with the same
numbers such that $\Delta_0=\Delta_1$. Both operators can be considered
as the same operator $L_0$ acting on the functions on the Lattice $Z$.

\begin{eqnarray}
(L_0\psi)_n=\psi_{n-1}+\psi_{n+1}, n\in Z\\
-L_0=\Delta_0+2=\Delta_1+2\nonumber
\end{eqnarray}

Two standard bases of solutions for
 the equation $L_0\psi=\lambda\psi$ can be given by the
obvious formulas

\begin{eqnarray}
\psi^{\pm}_n=a_{\pm}^n,a_{\pm}=1/2(\lambda\pm \sqrt{\lambda^2-4})\\
C_n=\frac{\psi_n^-a_+-\psi_n^+a_-}{a_+-a_-},S_n=\frac{\psi_n^+-\psi_n^-}
{a_+-a_-}\nonumber\\
C_0=1,C_1=0,S_0=0,S_1=1\nonumber\\
\psi^{\pm}=C+a_{\pm}S
\end{eqnarray}
\end{exa}

More general operators $L$ on the Lattice $Z$ appeared in
 the discretized Theory of Solitons (theory of Toda Lattice)
 beginning in 1974  (see in the survey  and encyclopedia articles
 \cite{DMN76,Encycl-4}). They have form
 of the second order selfadjoint Schrodinger Operators on the graph
 $\Gamma=R^1$ above
 \begin{eqnarray}
 (L\psi)_n=c_{n-1}\psi_{n-1}+V_n\psi_n+c_n\psi_{n+1}\\
 c_n=b_{n:n+1}=d_{n:n+1}, \nonumber 
 \end{eqnarray}

Solving the most fundamental problems of Spectral Theory for these operators
 for rapidly decreasing
coefficients $c_n-1, V_n$ and 
 for periodic  (quasiperiodic)
coefficients,  the so-called {\bf Wronskian} for any pair
of solutions has been used.

Our goal is to invent and to use a natural analog of this quantity for
any finite order Schrodinger Operator on the arbitrary Graph $\Gamma$
satisfying to 
the conditions above.

Consider any pair of solutions $L\psi=\lambda\psi, L\phi=\lambda\phi$
on the Graph $\Gamma$.

\begin{deff}

The following quantity will be called {\bf Wronskian} for any pair of solutions.

For vertices:
\begin{eqnarray}
W=\sum_R W_R\nonumber         \\
W_R(\phi,\psi)=b_{P:P'}(\phi_P\psi_{P'}-\psi_{P'}\phi_P)\\
P\bigcup P'=\partial R\nonumber
\end{eqnarray}

For Edges:
\begin{eqnarray}
W=\sum_R W_R \nonumber\\
W_R(\phi,\psi)=\sum_{R'}d_{R:R'}(\phi_R\psi_{R'}-\psi_R\phi_{R'})\\
R\bigcap R'=P\nonumber
\end{eqnarray}
\end{deff}

We can see that our formula for the Wronskian is very much the same
as for the line in the case of vertices, but for the edges it is less obvious.
 For example,
it contains  summation  along the edges $R'$ meeting our
edge $R$ in one point $P$ only. Second boundary point $P'$ where
$\partial R=P\cup P'$  does not appear in the sum. Therefore,
in this case  the correctness of our definition should be proved.

\newtheorem{theo}{Theorem}
\begin{theo}

For any pair of solutions $\phi,\psi$ of the second order difference
equation $L\phi=\lambda\phi,
L\psi=\lambda\psi$, Wronskian is a well-defined 1-chain $W$
 on the Graph $\Gamma$ (i.e. a complex-valued
function of the oriented edges $R=PP'$) whose boundary is equal to zero
\begin{eqnarray}
\partial W=0
\end{eqnarray}
Therefore our Wronskian belongs to the first homology group
$H=H_1^{open}(\Gamma, Z)$ modulo infinity (if Graph is noncompact).
It defines an $H$-valued skew symmetric scalar product
on the spaces of solutions.

\end{theo}

Proof. In the case of vertices, Wronskian is obviously well-defined as
1-chain. From the equality $(L\phi)_P\psi_P-(L\psi)_P\phi_P=0$
we extract immediately that
$$\sum_{P'\cup P=\partial R'}W_{R'}=0$$

Here  the edges  are taken with such orientation that they end in $P$.
However, this equality means precisely that $\partial W=0$ by definition
of the boundary operator.

Consider now the case of edges. Our formula above
defines correctly this quantity as $W_{R,P}$ depending on the edge $R$
and its vertex $P$. Starting from the same equality
$(L\phi)_R\psi_R-(L\psi)_R\phi_R=0$ for the pair of solutions,
we can see that the left-hand part is obviously equal to the sum
of two expressions. One of them is precisely  equal to $W_{R,P}$
as it was defined above for the case of edges.
The second one is equal to the
analogous sum $W_{R,P'}$ with point $P$ replaced by $P'$. So we have
\begin{eqnarray}
W_{R,P}+W_{R,P'}=0
\end{eqnarray}

Therefore we have $W_{R,P}=-W_{R,P'}$. Wronskian is well-defined
as a 1-chain. Consider now the boundary of it.
\begin{eqnarray}
(\partial W)_P=\sum_{P\in R}W_R=\\
=\sum_{R'\neq R}\sum_R d_{R:R'}(\phi'\psi-\psi'\phi)\nonumber\\
R'\bigcap R=P, \phi'=\phi_{R'},\phi=\phi_R\nonumber
\end{eqnarray}

However, in the last sum any fixed pair $R,R'$ appears twice
with opposite signs. Therefore the total sum is equal to zero.
Our theorem is proved. \footnote {It is strange , but
 the present author was not able to find this elementary fact
 in the literature (I asked several experts in the theory of graphs
 and operators on them). It does not surprise me for the edges, but
I cannot believe that this fact is new for the case of vertices.}

{\bf Higher Order Operators:}
For the case of higher order Schrodinger operators acting on vertices,
we define Wronskian in the same way as for the second order operators.
For any pair of interacting vertices $P,P'$, we fix one simple path
\begin{eqnarray}
I_{PP'}=R_1,R_2,\ldots,R_k, \partial I=P\cup P'
\end{eqnarray}

Elementary 1-chain associated with interacting
pair $PP'$ is defined as before. Full Wronskian is a sum of these
expressions along all interacting pairs of vertices:
\begin{eqnarray}
W=\sum_I W_I, W_I=b_{P:P'}(\phi_{P'}\psi_P-\psi_{P'}\phi_P)\\
\partial W=0\nonumber
\end{eqnarray}

As before, it is easy to prove that
$$W\in H_1^{open}(\Gamma, Z)$$

For the complex solutions $\psi_P$, one may consider Wronskian
$W(\psi,\bar{\psi})$ as a {\bf Quantum Current};  its Topological property
to be  1-cycle is a {\bf Kirchhof Law}.

\newtheorem{rema}{Remark}

{\bf Multidimensional Simplicial Complexes:}
For the natural classes of  Schrodinger operators acting on the spaces of
$k$-chains in Simplicial Complexes $K$ with $k>1$, we can define analogous
quantity (Wronskian) as a function on the set of all
 pairs $W_{S_k,S_{k-1}}$
(here we have a $k$-simplex and  its $k-1$-face). It is exactly the
class of operators
for which $k$-simplices can interact if they have common ($k-1$)-face.
This is an exact definition of this class of operators and of the Wronskian:
\begin{eqnarray}
(L\psi)_{S_k}=\sum_{S_{k-1}\in S_k}\sum_{S_{k}'\cap S_k=S_{k-1}}b_{S_{k}':S_k}\psi_{S_{k}'}\\ 
W_{S_k,S_{k-1}}(\phi,\psi)=\sum_{S_{k}'\cap S_k=S_{k-1}}
b_{S_k:S_{k-1}}(\phi'\psi-\psi'\phi)\\
\psi'=\psi_{S_{k'}},\psi=\psi_{S_k}\nonumber
\end{eqnarray}

The following theorem is true.

\begin{theo}

1.The full sum
of ''Wronskians'' along
all $k-1$-faces of every $k$--simplex is equal to zero;

2.  The full sum of ''Wronskians'' along all $k$-simplices with common
$k-1$-face is also equal to zero.
\end{theo}

Proof of this theorem is exactly the same as for $k=1$.

\begin{rema}

For  nontrivial applications of the theorems above we need to consider
such classes of operators that for some $\lambda$ the space of solutions is
at least 2-dimensional, otherwise our Wronskian will be identically
 equal to zero. There are two natural sources for that for the graphs:

 1.Graph $\Gamma$ and operator $L$ admit nontrivial symmetry group.

 2.Graph $\Gamma$ has several number of ''Tails'' at infinity
 and operator $L$ has  asymptotically  constant coefficients.
 It is exactly a case for which Scattering Theory can be developed.
  We shall do this in the next paragraph. In particular,
   we shall demonstrate
  that ''Unitarity'' of Scattering follows from the Topological Property
  of Wronskians established above.
 \end{rema}

\vspace{1cm}

\centerline{\Large Scattering Theory for  Graphs with Tails }
\centerline{and}
\centerline{\Large Symplectic Geometry}

\vspace{0.3cm}
We consider graphs with $k$ tails here. It means precisely that
outside of the finite domain this Graph is 
 isomorphic to the union of $k$ positive half
lines  (''Tails'') $z_1,z_2\ldots,z_k$.
Let us choose some initial vertices $P_j,j=1,2,\ldots,k$
in these tails and attribute to them the number of the tail and
zero. Other vertices in the same tail behind $P_j$
will be naturally numerated by
the number of the tail and
positive integers $n\in Z^+$. We attribute to the edge $R=(n,n-1)$
number $n$, as in paragraph 1 for the line, $n=0,1,2,\ldots $.

Any Graph $\Gamma$ with $k$ tails can be naturally represented in unique way
as
a union of the finite subgraph without ends $\Gamma'\subset \Gamma$
  with some number
of Trees attached to it in the vertices $Q_l,l=1,\ldots, s$.
Obviously we have $s\leq k$ because every tree contains at least one tail
(maybe more).

\begin{deff}

We call Graph $\Gamma'$ the {\bf Basis} of our Graph $\Gamma$. Points
$Q_l$
from which Trees grow up in the Graph $\Gamma$ will be  called {\bf Nests}.
Several trees can grow up from one nest.
Connected Graph $\Gamma$ is Topologically trivial iff its basis $\Gamma'$ is
equal to one point. We call graphs $\Gamma$ with $k$ tails
{\bf Diagrams} for which Scattering Processes naturally can be defined.
\end{deff}

Consider now a class of second order Schrodinger Operators $L$ on 
Graph $\Gamma$ with $k$ tails acting on the vertices or on
the edges such that outside of the finite domain we have in the tails
for $n\geq n_0$ and all $j=1,\ldots,k$:
\begin{eqnarray}
-L=-L_0=\Delta_i+2, i=0,1
\end{eqnarray}

Outside of this finite ''domain of interaction'' our equation
$L\psi=\lambda\psi$
has solutions in the tails
$\psi^{\pm}_{jn}$ and $C_{jn},S_{jn}$ for $j=
1,\ldots,k$ and $n>n_0$, described above (see formulas  (6,7) in paragraph 1).

\begin{deff}

Symplectic space $H^{2k}$ with basis $C_1,S_1,\ldots,C_k,S_k$ and
real-valued skew scalar product such that
\begin{eqnarray}
<C_j,C_p>=<S_j,S_p>=0\nonumber \\
<C_j,S_p>=\delta_{jp},j,p,=1,\ldots,k
\end{eqnarray}
will be called an {\bf Asymptotic Symplectic Space}.

\end{deff}
Solutions $\psi^{\pm}_j$ in the tails
can be expressed as  complex linear combinations
in the same basis with natural linear extension of the same scalar product.

\begin{deff}

 Vector $\psi^{ass} \in H^{2k}$ such that there is a continuation of it as
 a solution $\psi$ on the whole Graph $\Gamma$
 will be called an {\bf Asymptotic Vector} for the operator $L$;
$\lambda$-dependent linear space $T_{\lambda}$
 of all asymptotic vectors $\psi^{ass}\in T_{\lambda}$ for the operator $L$
 will be called  {\bf Space of Symplectic Scattering Data} for $L$.
 \end{deff}

\begin{theo}

For any selfadjoint real second order Schrodinger Operator $L$,
the corresponding linear space of Symplectic Scattering Data  is
Lagrangian subspace in  the Asymptotic Symplectic Space 
for every complex value of $\lambda$.
\end{theo}

Proof. We shall demonstrate that this fact is, in fact, topological.
It follows from  Theorem 1  which claims that
Wronskians are homological cycles in our Graph for every pair of solutions
$\phi,\psi$.

At the same time, we know the following elementary topological
properties of tails
in the connected graphs:

1.Any piece of tail $z_j$ may appear as
 a nonzero part of 1-cycle only if all edges
of this tail belong to this cycle with the same coefficient;

2.Individual tail $z_j$ never has a continuation to Graph
$\Gamma$ as a cycle
containing other tails with coefficients equal to zero. The
difference between
any pair of  tails $z_j-z_p$
can be continuated to Graph $\Gamma$ as a cycle such that
other tails do not participate in it.

For the Wronskian of two solutions on $\Gamma$ we have
\begin{eqnarray}
W(\phi,\psi)=\sum_{j=1}^{j=k}\alpha_jz_j+(finite)=\\
=\sum_{p=2}^{p=k}\beta_p(z_1-z_p)+(finite)\nonumber
\end{eqnarray}
From this equality we conclude that
\begin{eqnarray}
\sum_{j=1}^{j=k}\alpha_j=0
\end{eqnarray}

At the same time we know that this sum is exactly a skew symmetric
scalar product of asymptotic vectors:
\begin{eqnarray}
\sum_{j=1}^{j=k}\alpha_j=<\phi^{ass},\psi^{ass}>
\end{eqnarray}
Therefore our theorem is proved.

 Our Schrodinger Operators will be considered
now as a selfadjoint operators
in the Hilbert Spaces of square integrable functions $L_2^i(\Gamma)$
for two cases:  vertices $i=0$ and edges $i=1$
as before. The subset of $\delta$-
functions $\delta_P\in L_2^0$ or $\delta_R\in L^1_2$
contains functions equal to 1 on some vertex (edge), and
zero otherwise.

{\bf  For trivial reasons we always have a  continuous
spectrum with multiplicity equal to $k$ for $|\lambda |\leq 2$. Therefore
the results of  paragraph 1 are very useful here. }

 Sometimes we may have exceptional discrete
eigenvalues for $|\lambda |\leq 2$ drown in the continuous spectrum
(see below). A ''Normal'' discrete spectrum appears for $|\lambda|>2$

\begin{deff}
We call the
 {\bf  Scattering Zone of Spectrum} the area
  where $|\lambda |\leq 2, \lambda\in R$.
 The interval of real $|\lambda |>2$ we call
  the {\bf Zone of Normal Discreet Spectrum},
 for which every eigenfunction has exponential decay in all tails
 and is nonzero at least in one tail.  {\bf Exceptional Discrete
 Spectrum} may appear for any real $\lambda$. It is such that
 corresponding eigenfunctions are identically equal to zero in all tails.
 \end{deff}

 In the Scattering Zone we have
 \begin{eqnarray}
 \psi^{\pm}_{jn}=\exp\{\pm in\theta_j(\lambda)\}=a_{\pm}^n, n>n_0\\
 \theta(\lambda)\in R, |\lambda|\leq 2\nonumber
\end{eqnarray}

In the Zone of Normal Discrete Spectrum we have
$$\theta(\lambda)\in iR, i^2=-1$$

So in the last zone both solutions $\psi^{\pm}$ are real. One of them
has exponential decay for  large  $n$:
\begin{eqnarray}
\psi^+_j\rightarrow \infty, n\rightarrow\infty, j=1,\ldots,k\\
\psi^-_j\rightarrow 0, n\rightarrow\infty\nonumber
\end{eqnarray}

\noindent Let us denote the open halflines on the real $\lambda$ line
by
\begin{eqnarray}
I_+=(2<\lambda<\infty),I_-=(-2>\lambda>-\infty)
\end{eqnarray}

\newtheorem{corr}{Corollary}
\noindent Using the result of the previous Theorem we have a pair of mappings
\begin{eqnarray}
T^+:I_+\rightarrow \Lambda_k,T^-:I_-\rightarrow \Lambda_k
\end{eqnarray}

\noindent defined by the Lagrangian Planes $T_{\lambda}$ for  both cases
$\lambda\in I_{\pm}$. Here $\Lambda_k$ is a {\bf Lagrangian
Grassmanian} whose points are  Lagrangian $k$-planes in $H^{2k}$.
There is a canonical codimension 1 cycle (see \cite{Encycl-4})
$$Z\subset \Lambda_k$$
containing all Lagrangian planes with a nonempty intersection with a set of
directions  $\sum_j\kappa_j\psi^-_j$. You may say that
this cycle contains all singularities of the projection, when you project
 Lagrangian planes into the  linear span of all vectors
$\psi^+_j$  forgetting halfbasis $\psi^-_j$.

\begin{corr}

Normal discrete eigenvalue appears exactly where the Lagrangian Plane
$T_{\lambda}$ crosses the cycle $Z$. Therefore two ''Morse Indices''
are defined; they
characterize topological properties of normal spectra for $\lambda>2$
and $\lambda<2$: they are
the intersection indices of the curves $T^{\pm}$ with the cycle $Z$
defined above.
\end{corr}
We can see that decay of any eigenfunction belonging to the Hilbert
space should be exponential here (or it should be equal to zero in all tails--
see later).

\begin{rema}

I did not clarified yet whether these intersection indices are
exactly equal to
the numbers of eigenvalues (i.e. all crossing points for the generic case
 have sign +) or not.
\end{rema}
Let us define now  {\bf Scattering Matrix} for Operator $L$. Consider
the Scattering Zone $|\lambda|\leq 2$ where our basic complex solutions
$$\psi^{\pm}_{jn}=\exp\{\pm in\theta_j(\lambda)\}$$
are complex adjoint to each other.
 Take them as a basis for the complexified Symplectic
Space $H^{2k}_c$ where Lagrangian subspace $T_{\lambda}$ is given
by the theorem above. Generically this subspace can be interpreted
as a graph for the linear map from halfbasis $\psi^+=(\psi^+_j)$
 to the halfbasis
$\psi^-=(\psi^-_j)$
\begin{eqnarray}
\psi^+_p\rightarrow \sum_j s_{jp}\psi^-_j     \\
\psi^+\rightarrow S(\lambda)(\psi^-)\nonumber
\end{eqnarray}
This corresponds to the choice of basis in the form
\begin{eqnarray}
e_l=\psi^+_l+\sum_j s_{jl}\psi^-_j\in T_{\lambda}
\end{eqnarray}

\begin{corr}

In the Scattering Zone the Scattering Matrix $S_{\lambda}$
is always unitary $S\in U_k$ and symmetric $S^t=S$
\end{corr}
Proof.
For $k=2$, the relationship between Lagrangian planes and real
unimodular matrices
 was mentioned as an example in the elementary
  textbook of Arnold (see \cite{Atext}):
 From the Lagrangian plane in $H^4$, we come to the standard Monodromy Matrix
  $M\in SL_2(R)$ which maps basis $C_1,S_1$ into the basis
 $C_2,S_2$. This matrix was always  in use in Classical Math Literature
 for the second order Sturm-- Liouville operator on the line.
 Coming to the complex bases $\psi^{\pm}_j$ and $\psi^{\pm}_j$,j=1,2,
  we get a monodromy matrix in
 the group $SU_{1,1}$ which is isomorphic to $SL_2(R)$.
 Monodromy matrix is  unimodular because of wronskian property.
 Algebraic transform from this to the Scattering Matrix has the standard
  name of
 {\bf Caley Transformation} in this case. It was used by quantum physicists
 in the Scattering Theory for the Schrodinger Operators. {\bf Let us
  point out that standard
  Quantum Scattering Matrix $S'$ on the line differs from our $S$ by the
 multiplication on permutation matrix $P$ from one side:
 $S'=SP$.} For example,
 diagonal elements of $S$ are equal to the Reflexion Coefficient,
   not to Transmission Coefficient as in standard matrix $S'$.
 The famous Reflectionless Operators have Antidiagonal Scattering Matrix $S$
 in our sence for $k=2$.
 For any  $k$ we immediately deduce from lagrangian property
 that
 $$<\psi^+_j+\sum_l s_{jl}\psi^-_l,\psi^+_p+\sum_qs_{pq}\psi^-_q>=
(s_{jp}-s_{pj})<\psi^+,\psi^->=0$$

$$<\psi^+,\psi^->=\sqrt{\lambda^2-4}\neq 0$$

Therefore our Scattering Matrix is symmetric.

 To prove unitarity we need
to use reality of the Lagrangian plane. Consider now a real basis in
the plane $\Lambda$:
\begin{eqnarray}
(\sum_lt_{jl}e_l)=
T(\psi^+)+TS(\psi^-)
\end{eqnarray}

Its adjoint has a form
$$\bar{T}\bar{S}(\psi^+)+\bar{T}(\psi^-)\in \Lambda_k$$
$$\bar{\psi^+}=\psi^-$$

From that we conclude
$$\bar{S}^{-1}\bar{T}^{-1}\bar{T}=S=S^t$$
because we are coming to the same basis $e\in \Lambda$.
This line proves unitarity of $S$.
Our Corollary is proved.

\begin{rema}
We can start with real basis in $T_{\lambda}$ of the form
$$\bar{A}(\Psi^+)+A(\Psi^-)$$
From the same arguments as above we deduce that $A$ is a unitary matrix,
and
$$S=AA^t\in U_k$$
This is an imbedding of the space $\Lambda_k$ in $U_k$ as a set of all
symmetric unitary matrices.
\end{rema}

Let us describe now an Exceptional Discrete Spectrum.
The following simple theorem is true.
\begin{theo}

Exceptional Discrete Eigenfunctions are completely defined by the
eigenfunctions $\phi$ on the basis $\Gamma$'  equal to zero in all
nests (case of vertices).
\begin{eqnarray}
L'\phi=\lambda\phi (L'=D L D)\\
\phi(P_l)=0, l=1,\ldots,s\nonumber
\end{eqnarray}

Here $D^2=D$ is the projection operator from functions on
$\Gamma$ to the
functions on $\Gamma'$, putting all values outside subgraph equal to zero.
\end{theo}

\begin{rema}
In the case of edge operators we call 'edge-nest' any    edge $R$
 outside of $\Gamma$' touching some vertex-nest $P_l$. After that
 replacement
 all theorems containing the word 'nest' remain true for edge operators.
 \end{rema}

Proof of this theorem is very simple.
 Every such eigenfuction on the basis $\Gamma'$
can be continuated to $\Gamma$, taking zero value in all tails.
This gives us an exact one-to-one correspondence between
exceptional eigenvalues in $\Gamma$ and such special eigenvalues
in $\Gamma'$, equal to zero in all nests.

\begin{corr}

The property of Operator $L$ to have at least one exceptional
eigenvalue has codimension not less than  the number of nests  or edge--nests
(in the space of all real selfadjoint Schrodinger Operators with finite
 domain where operator is different from the standard constant operator $L_0$).
 In
particular, it is always not less than one, and Generic Operators
have no exceptional spectra. 
 After generic small perturbation of $L$
all discrete eigenvalues with $|\lambda|\leq 2$ disappear.

\end{corr}

Proof of this corollary immediately follows from the fact that finite Graph
$\Gamma'$ has only a finite number of eigenvalues. Without the symmetry
group, corresponding eigenfunctions are generically nonzero in the nests
(edge-nests).

\begin{rema}
For the Graphs and operators with nontrivial symmetry group  this
simple counting of parameters does not work. Exceptional eigenvectors
may necessarily appear in some symmetric cases--see examples 2 and 3 below.
\end{rema}

\begin{rema}
As Misha Gromov often explains in his lectures,  Hyperbolic Geometry
is visible from the infinity as one-dimensional one. Therefore we
may conclude, that for the discrete groups in 2D Lobachevski Plane
with Noncompact Fundamental Domain of
finite volume, Spectral Theory  of the
Laplace-Beltrami Operator should look in a sence 'similar'
 to the one on the graphs with
$k$ tails. We discussed this analogy with D.Kazdan, who pointed out to me
that for arithmetic subgroups there are many discrete eigenvalues
drown in the continuous spectrum. They disappear after
nonarithmetic perturbation,
as Peter Sarnak pointed out.
In the case of graphs with k tails,
 we  have simplified version of this  picture for 
 operators with symmetry: exceptional eigenvalues
  disappear after generic nonsymmetric perturbation.
\end{rema}

\begin{exa}

Let $k=1$. Our Graph $\Gamma$ is equal to  finite subgraph $\Gamma$',
plus tail attached to it at one point (nest $P_1$).
For  generic graphs $\Gamma$, we have only Lagrangian plane $T_{\lambda}$,
which is one dimensional, and no exceptional eigenvalues.
In the area where the operator is free (i.e. in the tail $n>n_0$),
we have one real solution $\phi=a\psi^++b\psi^- $ for every real $\lambda$.
In the  Scattering Zone $|\lambda|\leq 2$, we have
\begin{eqnarray}
\bar{\phi}=\phi, \bar{\psi}^+=\psi^-, \bar{b}=a\\
|b/a|=1,\lambda\in [-2,2]\nonumber
\end{eqnarray}

By definition of the Scattering Matrix, we have here
$$s_(\lambda)=b/a\in U_1$$

For the Zone of Normal Discrete Spectrum $|\lambda|>2$,
we have an analytic continuation  of scattering coefficient
$s(\lambda)$
which has no sense of scattering anymore. Its poles $a=0$
give us points of normal discrete spectrum $\lambda_m$.

If exceptional spectrum drown in the continuous one
exists for this operator, we have
an eigenfunction $\phi'$ equal to zero in the tail.
So we have a two-dimensional eigenspace and can consider Wronskian
as a skew scalar product.

{\bf Wronskian $W(\phi,\phi')$ for this exceptional value of $\lambda$
is equal to some finite cycle $z\in H_1(\Gamma')$. For the graphs and
operators with $Z_2$-symmetry this possibility  is generic, i.e., in general
 it
cannot be  destroyed by perturbation.}

To illustrate the last statement, consider Graph $\Gamma$ with one tail
and triangle $\Gamma'=0AB$ attached to  the  vertex 0.
 Its vertices are numerated
by nonnegative integers and letters A,B, $P=...,n,\ldots,0,A,B$ with edges
$R_n=[n,n-1],R_A=[0A],R_B=[0B],R_{AB}=AB$.
 Take coefficients of the vertex Schrodinger
Operator $L$ in the form
$$v_n=0,n> 0,v_0=u,v_A=v,v_B=w, b_{n:n-1}=1, n>0, b_{0:A}=a,b_{0:B}=b,
b_{A:B}=c$$
We have following set of equations:
$$\left(\begin{array}{ccc}
u-\lambda&a&b\\a&v-\lambda&c\\c&b&w-\lambda\end{array}\right)
\left(\begin{array}{c}\psi_0\\ \psi_A\\ \psi_B\end{array}\right)=
\left(\begin{array}{c}-\psi_1\\0\\0\end{array}\right)$$
For the exceptional eigenvector we need such solution that
$\psi_1=\psi_0=0$. It leads to the condition:
$$\lambda_{ex}=w-bca^{-1}=v-acb^{-1}$$
If $|\lambda_{ex}|<2$, we have an exceptional eigenvalue drown
in the continuous spectrum.

$Z_2$--symmetry leads to the following coefficients:
$$v=w, a=b$$
Here we have $\lambda_{ex}=c$.

 \end{exa}

\begin{exa}

Let us consider vertex operators for $k=2$.
 We have two tails here and one or
two nests in the Graph $\Gamma'$.
For almost all $\lambda$ we have here  well-defined Monodromy
Matrix as on the line. However, on the line monodromy was well-defined
for all $\lambda$.

Here we may have isolated ''singular'' values of $\lambda$
for which Lagrangian plane $T_{\lambda}$ is special: a  monodromy
 map from the free basis of one tail into free basis of the
 another tail does not exist.
It  happens if our Lagrangian plane has a basis $\phi_1,\phi_2$
such that $\phi_1=0$ in the second tail and $\phi_2=0 $ in the first tail.

{\bf Wronskian $W(\phi_1,\phi_2)$ is  equal
to some finite cycle in $\Gamma'$ for such 'singular' $\lambda=\lambda^*$.}

 It might happen even in the Scattering Zone as a generic possibility.
 Such singular values $\lambda=\lambda^*$ may be real, or they may appear by the
 complex adjoint pairs.

 To illustrate the last statement, we consider two cases:

 1.$\Gamma'$ is a triangle $[0AB]$ where the vertex $0$ is a nest for
 both tails.

 2.$\Gamma'$ is a triangle $[0_10_2A]$ where the vertices $0_1,0_2$
 are the nests.

In the Case 1 we have vertices
$$P=...,n_1,\ldots,0_1=0,\ldots,n_2,\ldots,0_2=0,A,B$$
and coefficients
$$b_{n_i:n_i-1}=1,n_i>0,i=1,2,b_{0:A}=a,b_{0:B}=b,b_{A:B}=c,
v_0=w,v_A=u,v_B=v$$
As elementary calculation shows, we have solution $\psi_i$
equal to zero in
the  tail $i=1$ or $i=2$      iff
 $$(u-\lambda^*)(v-\lambda^*)=c^2$$
Here we have real  $\lambda^*$ only, and $W(\psi_1,\psi_2)=0$

In the Case 2 we have vertices
$$P=...,n_1,\ldots,0_1,\ldots,n_2,\ldots,0_2,A,n_i\geq 0$$
 and coefficients
 $$b_{n_1:n_1-1}=b_{n_2:n_2-1}=1, n_i>0,b_{0_1:0_2}=a,
 b_{0_1:A}=b,b_{0_2:A}=c,u=v_{0_1},v=v_{0_2}, w=v_A$$
 After elementary calculations we have
 $$\lambda^*=bca^{-1},W(\phi_1,\phi_2)=bca^{-1}[0_10_2A]\in H_2(\Gamma,R)$$
 Here $\phi_i=0$ in the tail with number $i, i=1,2$, and $\phi_i(A)=1$.
 Exceptional eigenvalues with eigenfunctions
 equal to zero in both tails do not exist for this
 simple graph.

 \end{exa}

For the cases $k\geq 3$ the  situation is much more complicated.

\noindent Let us point out here that a very simple argument
leads to the existence of  discrete spectrum $\lambda_q>4$
for Laplace-Beltrami operators.
Apply operator $L$ to the set of $\delta$-functions. Consider
the square of norm
$$(L\psi,L\psi)=||L\psi||^2$$
and the maximum of it for all $\delta$-functions. We denote this maximum by
$M_L$.
For the Laplace-Beltrami operators,  we shall use $L=\Delta+2$
as before. This choice of constant is optimal for this estimate.
{\bf  If this quantity is bigger than 4,
we  conclude that discrete spectrum exists for $L$.}
 Moreover, this estimate is
certainly nonexact. So, discrete eigenvalue $\lambda_q>2$ exists also
if our
estimate is exactly equal to 4 on the set of $\delta$-functions.
We are coming finally to the following estimates sufficient for the existence
of discrete spectrum in  most cases:

1.Vertices
$$M_L=max_P(\sum_{P'}b_{P:P'}^2+V_P^2)\geq 4$$

\noindent For the vertex Laplace-Beltrami, we have $M_L=max_P\{m_P+(m_P-2)^2\}$

2.Edges
$$M_L=max_R(\sum_{R'}d_{R:R'}^2+V_R^2)\geq 4$$

\noindent For the edge Laplace-Beltrami, we have $M_L=max_R\{m_P-1+m_{P'}-1\}$
where $P\cup P'=R$.
It is easy to improve last inequalities, replacing the Operator $L$
by $L^2$. Sasha Veselov obtained better estimates for the case of vertices,
 for example, as he privately informed me.

\vspace{1cm}
\centerline{\Large Factorization of Schrodinger Operators on Graphs}

\vspace{0.3cm}

\begin{deff}

We call Schrodinger Operator $L$ acting on vertices or edges
{\bf factorizable}
if it can be represented in the form

\begin{eqnarray}
L+C=QQ^+, C=const\\
L+C=Q^+Q+U_R\nonumber
\end{eqnarray}
\noindent Factorization is {\bf special} if function $U_R$ is equal
to constant. For the case of vertices we consider only
 special factorization.
Here operators $Q,Q^+$ are real adjoint to each other. An Operator
$Q+$ maps functions of vertices into functions of edges by the following
formula
\begin{eqnarray}
(Q^+\psi)_R=\sum_Pc_{R:P}\psi_P\\
(Q\psi)_P=\sum_Rc_{R:P}\phi_R
\end{eqnarray}
We call factorization {\bf formal} if coefficients of operators $Q,Q^+$
are not real (and they are not adjoint actually to each other).

\end{deff}

Elementary substitution of this into our equations leads to the
following result:

\begin{theo}
Representation of Operator $L$ in the factorized form is equivalent to the
set of equations:

1.For vertices
\begin{eqnarray}
b_{P:P'}=c_{R:P}c_{R:P'}, P\neq P'\\
V_{P}+C=\sum_{P\in R}c_{R:P}^2  \nonumber
\end{eqnarray}

2.For edges
\begin{eqnarray}
d_{R:R'}=c_{R:P}c_{R':P}, R\cap R'=P\\
V_R+C=c_{R:P}^2+c_{R:P'}^2+U_R\nonumber
\end{eqnarray}
\end{theo}

\begin{corr}

For Operators $L$ acting on edges: part of our factorization equations
 for the quantities $c_{R:P}$ through $d_{R:P}$ provides a complete set
 of algebraic equations for any given vertex $P$. {\bf This set of equations
 is overdetermined
 for such $P$ that $m_P>3$. Compatibility conditions (in the form of
 algebraic  constraint for the coefficients of operators)
 should be satisfied for
 factorization in this case.} If all coefficients $d_{R:P}$ are positive
 for  all closed neighbors $R\neq R'$, the solution $c_{R:P}^2$
 is also positive.
There is a formal factorization $L=QQ^t+U_P$ such that the
squares of coefficients $c_{R:P}^2$ are uniquely defined by the equation of
factorization above.
\end{corr}
\begin{corr}
For operators acting on vertices:
let all coefficients $b_{P:P'}$
are strictly positive for all closest neighbors $P\neq P'$.
 Take any finite contractible (''tree-like'')
subgraph $\Gamma''$
in the Graph $\Gamma$
and its initial vertex (nest) $P_0\in \Gamma''$. 
Take any value of the constant $C$ and any function $c_{R:P}$ given
 along the boundary of $\Gamma''$
 except the point $P_0$, where edges $R$ look inside
 of our selected subgraph from the boundary vertices $P$. There is a
 special formal
 factorization of $L$  with 
   functions $c_{R:P}^2$ uniquely defined by this data in the
   subgraph $\Gamma''$. There is a value of the constant $C$ depending
   on the subgraph $\Gamma''$ such that
   all quantities $c^2_{R:P}$ are positive.

   \end{corr}

\noindent   Proof of this  easy follows from the form of the factorization
    equations: we solve the equations above for the quantities
    $c_{R:P}^2$ in both cases. Let me point out that factorization
    is purely local in the case of edges.
     So we proved our corollary for the case of edges.

    Consider the case of vertices more carefully.
     From the data on the boundary of subgraph $\Gamma''$ we can find
    unique real solution for the squares  $c^2_{R:P}$ for any real constant $C$.
    However these quantities might take negative value. This leads to 
    complex solutions in terms of $c_{R:P}$. After that we find signs
    of $c_{R:P}$
    from the very simple part of our equation which does not contain
    squares, using initial data.

To prove the last part of the Corollary, we need to take
constant $C$ large enough and special initial data on the boundary of
 subgraph $\Gamma''$. We take large enough values of $c^2_{R:P}$
on the boundary after choosing  a large constant $C$. Here the edges $R$
are attached to the boundary points $P$ from inside of $\Gamma''$.
Solving the factorization equation in the direction to the endpoint $P_0$
we shall go through the number of steps (layers) in $\Gamma''$,
such that $c^2_{R:P}$ will be small,
 of the order $C^{-1}$ for the edges looking
from outside to the layer vertices. After that we shall find that
the values of $c^2_{R':P}$ will be large and closed to $C$ for the edges
looking inside of layers. This ansatz is selfconsistent. Therefore we
are coming to the desired positive solution.
Corollary is proved.

{\bf Conjecture}: Let coefficients  $d_{R:R'}$ and   their inverse $d^{-1}_{R:R'}$
for $P\neq P',R\neq R'$ are positive and bounded.  Let Graph $\Gamma$
is contractible and all numbers
 $m_P$ are also bounded.
There is some positive  value of $C$ that Operator $L$
admits a real factorization.

\noindent  Probably, this statement follows from the little inprovement
 of the same arguments as  last Corollary.

Until now we did not investigate factorization for the graphs with
nontrivial topology.

\vspace{1cm}
\centerline{\Large Appendix: Two Remarks}
\vspace{0.2cm}
\centerline{\Large  1.Nonlinear equations. 2. Fermionic Quadratic Forms}
\vspace{0.3cm}
We shall discuss here the continuous analog of our constructions
 for the case of vertices     and nonlinear generalizations.
 After that we describe some properties of real fermionic quadratic forms.

1. Consider any smooth manifold $M$ with Riemannian metric
 and the linear selfadjoint Schrodinger Operator $L$
acting  in the space of the scalar functions.
This operator can be obtained from  the
variational principle
$$S\{\psi\}=\int_M(\sum a^{jk}(x)\psi_k\psi_j+U(x)|\psi|^2)
\sqrt{ g(x)}d^nx$$
$$ \psi_k=\partial_k\psi$$

\noindent in the Hilbert space of square integrable functions,
where $g(x)$ is determinant of the Riemann tensor in the local
coordinate system $x^1,\ldots,x^n$. What Gelfand told me in 1971 is
that the expression
$$\int_B ((L\psi)\phi-(L\phi)\psi)\sqrt{g(x)}d^nx$$
\noindent for the arbitrary pair of functions $\psi,\phi$
is nonzero for the domain $B$. Using  the Stokes formula, it can be
reduced to the boundary $\partial B= D$. It   leads to the integral
$$\int_B((L\psi)\phi-(L\phi)\psi)\sqrt{g(x)}d^nx=\int_{D}W(\phi,\psi)d
\gamma$$
\noindent Here $d\gamma$ means corresponding area element on the boundary.
As we can see, what we get along the boundary is in fact
an analog of Wronskian in our terminology. For the selfadjoint extension of
operator $L$ in the domain $B$ you need to take  boundary conditions
in the form of some Lagrangian Plane
 in the last
 space of  pairs $(\psi,\phi)\in H$ of
 functions on the boundary.

 In the main part of this work, we defined and used
  a discrete analog (for the Graph $\Gamma$)
 of the quantity $W$ which is  now the density
 of vector field $W^j(x)\sqrt{g(x)}d^nx$ on the manifold $M$ in the
 continuous case. Our scheme corresponds in this case
 to the following:
 we take any 
 1-form $\omega_i(x)dx^i$ with compact support on $M$;
  $W$ can be considered  as a 1-chain on $M$ (''current'')
$$<z,\omega>=\int_M\omega_i(x)W^i(x)\sqrt{g}d^nx$$
\noindent For the pair of solutions $(\psi,\phi)$
of the equation $L\psi=\lambda\psi$, we are coming to the 1-cycle
associated with current $W$.

We did not clarify any continuous analogs of the cases $k>0$.

After the very useful discussion of our first work $\cite{N97}$
with  A.Schwarz in Maryland in December 1997,
we came to the conclusion that {\bf this
  construction has a natural generalization to
the nonlinear case: we may construct a closed H-valued 2-form on the space
of solutions for  nonlinear variational problems in both cases--
continuous and discrete, i.e., on $M$ and on the Graphs
$\Gamma$. Here $H=H_1^{open}(\Gamma,Z)$ for  graphs,
 and $H$ is a divergionless
current for manifolds}.  Construction of this 2-form
immediately follows
 from this work, applying  it to
 the operator of second variation along the solution
 $f$, to the pair of  variations for $\lambda=0$:
 $$W_f(\delta\psi,\delta\phi)=\Omega$$
\noindent This form is closed for $\lambda=0$.
 I will publish details in the next
work.

\begin{rema}

In one--dimensional case there is a natural class of continuous
 operators--the Schrodinger Operators on Graphs such
that in every edge $R_i$ a self--adjoint second order standard Schrodinger
Operator $L_i$ is given, characterized by the set of real potentials
$v_i(x), x\in R_i$
given in all  edges $R$ as the functions continuous in the closed edge
(i.e. including boundary).
 For any vertex $P$ we have
$k=m_P$ edges $R_i$ ending in it. Consider Symplectic Space $R^{2k}_P$
given as a direct sum of 2-spaces $R^2_i$ associated with every edge $R_i$
$$R^{2k}_P=R^2\oplus\ldots \oplus R^2$$
 with canonical coordinates $p_i,q_i\in R^2_i$.
 For any solution $L_i\psi_i(x)=\lambda\psi_i(x)$ in the edge $R_i$,
 we have  boundary values $\psi(P),\psi'(P)\in R_i^2$.
 Let for any vertex $P$ a Lagrangian Plane $\Lambda_P\subset R^{2k}_P$
is given. We call set of solutions $L_i\psi_i(x)=\lambda\psi_i(x), x\in R_i$
{\bf  solution of the real selfadjoint operator L on the graph
iff in any vertex $P$ full set of the boundary values belongs to the
Lagrangian Planes $\Lambda_P$}.

We may replace operators $L_R$  by the Monodromy Matrices $M_R\in SL_2(R)$:
$$M_R:R^2_P\rightarrow R^2_{P'}, \partial R=P\cup P'$$
All set of vertices can be considered as some kind of 'boundary'
 for the Schrodinger Operator $L=\cup L_R$ (union along all edges).
  The whole set of
 Lagrangian Planes $\Lambda=\cup\Lambda_P$ is some sort of 'selfadjoined
 extension' for the Operator $L$.  As P.Kuchment informed me in May 1998,
  these operators
 have been considered by mathematical and theoretical physicists
 for some concrete problems of Solid State Physics  and Superconductivity
 in the late 80-s, who formulated  self-adjoint junction condition in vertices in
 in the classical von Neumann-Krein form. The most popular
 conditions are  following:
 $$1.\psi_i(P)=\psi_j(P)=\psi(P), \sum_i\psi'(P)=\alpha\psi (P)$$
 $$2.\psi'_i(P)=\psi'_j(P)=\psi'(P),\sum\psi_i(P)=\alpha\psi' (P)$$
 Here  indices $i,j$ as before numerate
  all edges coming to the vertex $P$.
 (see \cite{ES})
  B.Pavlov with N.Gerasimenko
 obtained also some results
 in the Scattering Theory for such operators (see \cite{GP}). We shall compare
 their results with our ideas in the next work.
In the works of  P.Kuchment and A.Figotin
 a pseudodifferential model on graphs
 was developed for photonic crystalls (see for example \cite{FK}).
   Certainly, no one of them discussed
 such  problems in terms of Symplectic Geometry.

Consider now any subgraph $\Gamma_1\subset\Gamma$
 such that $\partial R\subset
\Gamma_1$ implies $R\subset\Gamma$. Therefore the subgraph $\Gamma_1$
attached to the other part of the Graph $\Gamma$ by some edges
$R_1,\ldots,R_p$ ending in the vertices $P_i\in\Gamma_1$ and in the vertices
$P'_i\in
(\Gamma minus \Gamma')$.
 For any selfadjoint real Schrodinger Operator $L$ on $\Gamma$  we can define
 naturally the $\lambda$-dependent Lagrangian Plane $\Lambda_{\Gamma_1}$
 in the Symplectic Space $R^{2p}=R^2_{P_1}\oplus\ldots\oplus R^2_{P_p}$
 generated by the values of the solutions $L_i\psi_i=\lambda\psi_i$
  along the edges $R_i$
 and their first derivatives 
 in the vertices $P_i$. All interaction of this subgraph with other part
 depends on this Lagrangian Plane. In a sence we may consider such
 subgraph equipped by the Lagrangian Plane as some more complicated kind of
 {\bf Selfinteracting Vertex} with $\lambda$-dependent boundary conditions.

 \end{rema}
2. {\bf Fermionic Quadratic Forms.}
Let us consider now a finite dimensional real linear space $R^n$
with basis $e_j$ and total space of external powers
$$\Lambda^{\star}(R^n)=\sum_{k=0}^{k=n}\Lambda^k(R^n)$$
\noindent Let us associate with every basic vector $e_j$ a Fermionic Creation
Operator $a_j$; we associate a vacuum vector $\eta$ with
 unity $1\in R=\Lambda^0\subset\Lambda^{\star}(R^n)$:

$$e_j=a_j(\eta)$$
 The total space of exterior powers has a natural basis
$$a_{j_1}\ldots a_{j_k}(\eta),j_1<\ldots <j_k$$
 for all $j,k$.

\begin{deff} A Real Fermionic  Quadratic Form is a selfadjoint operator $L$
with real coefficients acting  on the total space of exterior powers,
written in the form:
$$L=A_{pq}a_pa_q+B_{pq}a^*_pa_q+ C_{pq}a^*_pa^*_q+const$$
\end{deff}

 Here the Annihilation Operators $a^*_p$
 are adjoint to the Creation Operators
$a_p$. They  satisfy to the ''canonical'' relations
$$a_pa_q=-a_qa_p, a^*_pa_q+a_qa^*_p=\delta_{pq}$$
 We have obviously
$$A_{pq}=-A_{qp}, B_{pq}=B_{qp}, C_{pq}=-A_{pq}$$

\begin{theo}
Take the new (noncanonical) basis $a^{\pm}_j=a^*_j\pm a_j$ 
An Operator $L$ has the form
$$L=D_{pq}a^+_pa^-_q+const, D=A+B$$
\end{theo}

\begin{deff} A Bogolyubov Transformation or Canonical Transformation
 is an isomorphism of this Clifford Algebra
given by the change of canonical basis:
$$a_p=\sum P_{pq}b_q+\sum Q_{pq}b^*_q$$
$$a^*_p=\sum Q_{qp}b_q+\sum P_{qp}b^*_q$$
 Here  new basis $b^*_j,b_j$ satisfies to the same algabraic relations
as the original one (both of them should be canonical).
\end{deff}

\begin{theo}
For any Canonical Transformation both matrices $O_{\pm}=P\pm Q$
are  orthogonal  $O_{\pm}\in O_n(R)$. The Transformation Rule for the
matrix $D$ is following:
$$D=O_+D'O_-$$
where
$$L=D_{pq}a^+_pa^-_q=D'_{rs}b^+_rb^-_s$$

\end{theo}
This theorem can be easily proved by the direct elementary verification.
As a corollary from this we have

\begin{corr}
An operator $L$ can be reduced to the diagonal form using  Bogolyubov
Transformations above and  Transformation Rule for  matrix
$D=A+B$. Eigenvalues $\mu_{j_1j_2\ldots j_k}$ of the Operator $L$
 in the space $\Lambda^*(R^n)$ can be computed in the following way: 
$$\mu_{j_1}+\ldots\mu_{j_k}=\mu_{j_1\ldots j_k},j_1<\ldots <j_k$$
\noindent where $\mu_j $ are the eigenvalues of the 'Absolute Value Operator'
$$|L|=\sqrt{L^*L}$$
in the space $R^n$
\end{corr}

This algebraic statement has been found  as a lemma
in the present author's
Appendix to our joint work with Misha Shubin (see \cite{NS86}).
In this Appendix I  developed a nice analytic
Witten-like approach trying to find right analog of the Morse
Inequalities for the Vector Fields on  real manifolds: necessity
to diagonalize an  arbitrary
Real Fermionic Quadratic Form appears naturally here.

 (Let us remind that in the Witten's approach to
  the Morse inequalities for ordinary functions
we
have $A_{pq}=0$; therefore    only one orthogonal
 matrix has been used for  the
diagonalization).
Several experts in Quantum Field Theory pointed out to the present
author that this
 elementary observation has
 never appeared in the literature, so I decided to repeat it here
 once more.
  (The last one was A.Schwarz who told me that in the December 1997.)

\end{document}